\documentclass[aps, prl, reprint, floatfix, superscriptaddress, longbibliography]{revtex4-1}

\usepackage{graphicx}
\usepackage{amsmath}
\usepackage{epstopdf} 
\usepackage[colorlinks=false, pdfborder={0 0 0}]{hyperref} 

\newcommand{\reg}{\mathrm}
\newcommand{\didv}{\reg d I/ \reg d V_\reg{SD}}
\newcommand{\sm}{\sim\kern-.6ex}

\begin{document}

\title{Measurements of a Quantum Dot with an Impedance-Matching On-Chip LC Resonator at GHz Frequencies}

\author{M.-C. Harabula}
\author{T. Hasler}
\affiliation{Department of Physics, University of Basel, Klingelbergstrasse 82, CH-4056 Basel, Switzerland}

\author{G. F{\"ul\"op}}
\affiliation{Department of Physics, University of Basel, Klingelbergstrasse 82, CH-4056 Basel, Switzerland}

\author{M. Jung}
\affiliation{Department of Physics, University of Basel, Klingelbergstrasse 82, CH-4056 Basel, Switzerland}
\affiliation{
	DGIST Research Institute, DGIST, 333 TechnoJungang, Hyeongpung, Daegu 42988, Korea
}

\author{V. Ranjan}
\affiliation{Department of Physics, University of Basel, Klingelbergstrasse 82, CH-4056 Basel, Switzerland}
\affiliation{
	Quantronics Group, SPEC, CEA, CNRS, Universit\'e Paris-Saclay, CEA Saclay, F-91191 Gif-sur-Yvette, France
}

\author{C. Sch\"onenberger}
\affiliation{Department of Physics, University of Basel, Klingelbergstrasse 82, CH-4056 Basel, Switzerland}

\date{\today}

\begin{abstract}
We report the realization of a bonded-bridge on-chip superconducting coil and its use in impedance-matching a highly ohmic quantum dot (QD) to a 3\,GHz measurement setup. The coil, modeled as a lumped-element $LC$ resonator, is more compact and has a wider bandwidth than resonators based on coplanar transmission lines (e.g. $\lambda/4$ impedance transformers and stub tuners) at potentially better signal-to-noise ratios. 
In particular for measurements of radiation emitted by the device, such as shot noise, the 50$\times$ larger bandwidth reduces the time to acquire the spectral density. The resonance frequency, close to 3.25\,GHz, is three times higher than that of the one previously reported wire-bonded coil. 
As a proof of principle, we fabricated an $LC$ circuit that achieves impedance-matching to a $\rm{\sm 15\,k\Omega}$ load and validate it with a load defined by a carbon nanotube QD of which we measure the shot noise in the Coulomb blockade regime.

\end{abstract}

\maketitle

\section{Introduction}
Superconducting qubit control and readout of very resistive quantum devices need microwave (MW) resonators operating at frequencies in the range of $1-10$\,GHz. To efficiently measure radiation emitted from a device, one attempts to balance the high device impedance to the $Z_0 = 50$\,$\Omega$ characteristic impedance of the usual coaxial lines using an impedance-matching circuit. These circuits are often implemented with superconducting on-chip transmission lines. Examples include quarter-wavelength step transformer in the fluxonium qubit~\cite{Manucharyan2009}, stub tuners for quantum point contacts~\cite{Puebla-Hellmann2012} and QDs~\cite{Ranjan2015,thomas}. An alternative approach makes use of an $LC$ resonator built either from a lumped element inductor~\cite{Schoelkopf1998} or form on-chip coils~\cite{Chen2004}. However, the resonance frequencies have typically been limited to a maximum of $\sm 1$\,GHz~\cite{Yue1998,Xue2007,Fong2012}.
It is however important to improve $LC$ circuits, since they are much less demanding in chip footprint as compared to transmission line circuits. While a transmission line circuit at $5$\,GHz needs up to $\sm 1$\,cm in length, an appropriate on-chip coil can be fabricated on a $100\times$ smaller area. This is an important consideration for the future scaling of qubit networks for quantum processors. To understand the challenge, the key parameters of an $LC$ matching circuit are its resonance frequency given by $1/(2\pi\sqrt{LC})$ and its characteristic impedance $Z_{\rm c}$ defined as $\sqrt{L/C}$. In order to match to a high-impedance load, $Z_{\rm c}$ should be as large as possible, meaning that one has to aim for a large inductance. At the same time, the resonance frequency should stay high, constraining the capacitance to as small as possible values. In fact, the limiting factor in achievable resonance frequency is the residual stray capacitance in the circuit, which we minimize in our approach.

Addressing the compactness of MW resonators in this work, we fabricated a 200-$\mu$m-wide on-chip superconducting coil with a wire-bonded bridge (Fig.\,\ref{fig1}) and utilize it as a lumped $LC$ matching circuit in a carbon nanotube (CNT) QD noise experiment at a working frequency close to 3.25\,GHz. Compared to the one previously reported case of on-chip inductor with bonded bridge~\cite{Xue2007,Xue2010}, 
we achieved a threefold frequency increase with a similar footprint.
Comparable compactness has been achieved only with Josephson junction arrays acting as quarter-wavelength resonators~\cite{Altimiras2013,Stockklauser2017}.

\section{Sample fabrication}

\begin{figure}
\includegraphics{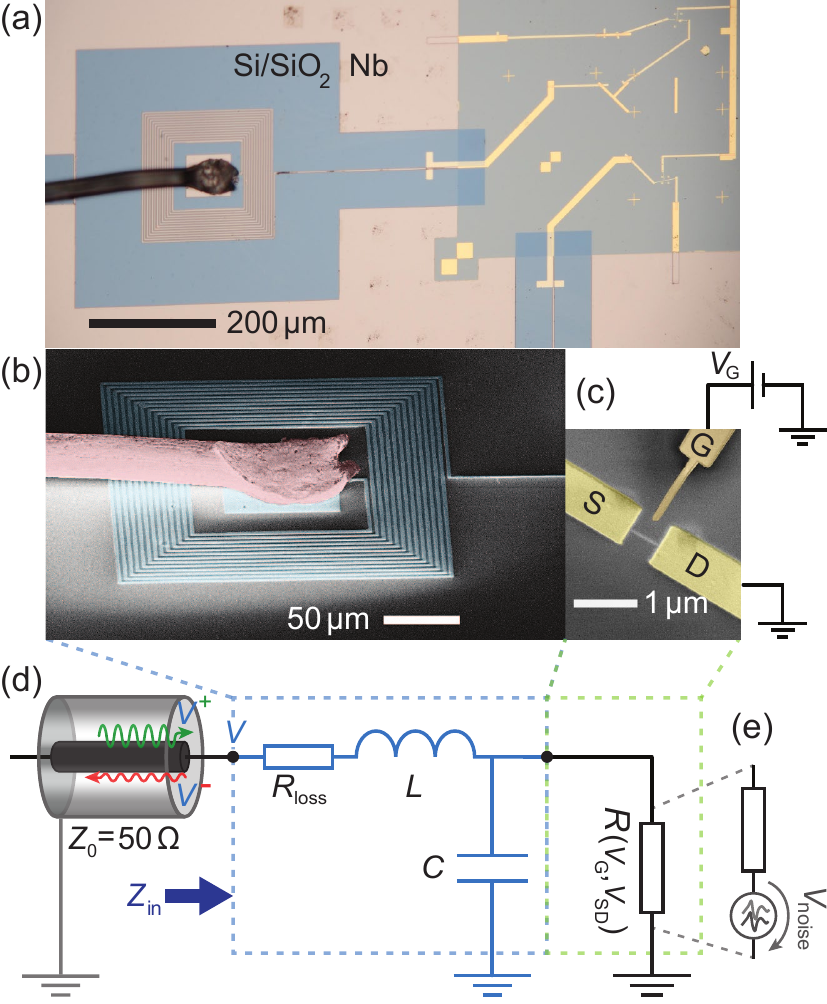}
\caption{
	\label{fig1} (a) Optical micrograph of the sample. (b) False-color SEM image of a superconducting Nb on-chip coil (cyan) connected via an Al bond wire (pink) to the measurement setup. The right end of the coil connects to a QD laying on the same Si/SiO$_2$ substrate. (c) False-color SEM image of the CNT QD with Ti/Au leads (S, D) and side gate (G). (d) Simplified sketch of the microwave-only electrical circuit: the coil is modeled as an $LC$ circuit, whose losses are caught by a resistance $R_\reg{loss}$ in series with the inductor $L$; the CNT is characterized by a resistance dependent on the gate and source-drain DC voltages; the $LC$ tank circuit partially matches $R$ to the $Z_0=50$\,$\Omega$ of the measurement setup line. In reflectance measurement the voltage $V=V^+ + V^-$ at the interface is the sum of the voltages given by the incident wave (green) and the reflected wave (red). (e) In noise measurements, there is no incident wave, $V^+ =0$. In contrast, a voltage source $V_\reg{noise}$ needs to be considered in series with the CNT resistance. This voltage is transmitted to become $V$ with a frequency dependent transmission function determined by the $LC$ network. 
}
\end{figure}

A planar spiral inductor raises the issue of connecting its inner end to the rest of the circuit. While it is possible to fabricate this bridge by lithography, we have realized that the close proximity of the bridge to the coil adds substantial capacitance, limiting the performance of the $LC$ circuit. In addition, the microfabrication of air-bridges of this kind add several fabrication steps. That is why we developed a solution of bonding a wire between a geometry-restricted inner pad and an external bigger pad at the low-impedance end.

The fabrication starts with preparing a region on the undoped Si\,/\,SiO$_2$ (500\,$\reg{\mu}$m\,/\,170\,nm) substrate where CNTs will be placed. Thus, we first evaporate Ti\,/\,Au (10\,nm\,/\,30\,nm) for markers and CNT partial contacts in a square area (top-right corner in Fig.\,\ref{fig1}(a)). We then protect it with a PMMA/HSQ bilayer resist. Afterwards, we sputter 100\,nm of Nb and lift the protection bilayer resist off. Subsequently, we e-beam-pattern bonding pads and the desired inductor in a new PMMA resist layer, then etch the Nb film with an $\reg{Ar/Cl_2}$ inductively coupled plasma; the surrounding Nb becomes the ground plane. Next, we stamp CNTs \cite{thomas,Viennot2014} in the predefined region, locate them using a scanning electron microscope (SEM), and contact the chosen CNT in one Ti/Au evaporation step to the coil and to ground (Fig.\,\ref{fig1}(a)), utilizing the partial contacts. In the same lithography step, a side gate is created at a distance of $\sm 300$\,nm from the CNT (Fig.\,\ref{fig1}(c)). With the device glued onto a sample holder, we use Al wire to bond the remaining end of the coil to a neighboring bigger pad. Due to the relatively small size of the inner pad ($\reg{70\times70\,{\mu}m^2}$, barely larger than the bonder wedge), the bonding operation is delicate (Fig\,~\ref{fig1}(b)). Finally, we connect the latter pad to the  MW line of the printed circuit board sample holder. Furthermore, the Nb ground plane of the sample is bonded with multiple wires along the wafer edge to the sample holder ground plane.

The square spiral inductor used in this experiment (Fig.\,\ref{fig1}(a,b)) has an outer dimension of 210\,$\mu$m, 14 turns with width $w = 2\,\mu\reg m$ and spacing $s = 2\,\mu\reg m$. In the device investigated here two of the turns are shorted, lowering the effective inductance and thus shifting up the resonance frequency by several percents.

\section{High-frequency setup and resonator characterization}
The simplified schematic of the MW circuit is illustrated in Fig.\,\ref{fig1}(d). The impedance-matching circuit, contained in the light-blue dashed rectangle, is modeled with lumped elements: $L$ is the inductance of the coil, $C$ its capacitance to ground, and $R_\reg{loss}$ the effective loss resistance, accounting for the RF (radio frequency) loss of the superconducting material and the dielectric loss in the substrate.
The input impedance $Z_\reg{in}$ (Fig.\,\ref{fig1}(d)) reads:
\begin{equation}
	\label{eq_Zin}
	Z_\reg{in} = R_\reg{loss}+\reg i \omega L+ \frac{1}{G+\reg i \omega C},
\end{equation}
with $G=1/R$, the load conductance.
$Z_\reg{in}$ can be approximated at $\omega_0 = 1/\sqrt{LC}$ by
\begin{equation}
	\label{eq_Zin_on_resonance}
	Z_\reg{in}(\omega_0) \approx R_\reg{loss}+\frac{Z_{\rm c}^2}{R}.
\end{equation}
This approximation is valid for $|Z_{\rm c}|\ll R$. Full matching is achieved at $\omega_0$ when $Z_\reg{in}(\omega_0)=Z_0$. Thus one obtains the condition $Z_{\rm c}=\sqrt{(Z_0-R_\reg{loss})R}$, meaning that the characteristic impedance of the $LC$ circuit should be equal to the geometric mean of $Z_0$ and $R$, if we neglect $R_\reg{loss}$. Typical values for $Z_{\rm c}$ are between one and a few k$\Omega$.

In reflectance measurements, a continuous sinusoidal wave is applied to the $LC$ circuit where it appears with amplitude $V^+$. Part of the incident signal is reflected back with amplitude $V^-$ due to impedance mismatch. The reflection coefficient $\Gamma$ is given by:
\begin{equation}
	\label{eq_Gamma}	
	\Gamma \equiv \frac{V^-}{V^+}= \frac{Z_\reg{in}-Z_0 } {{Z_\reg{in} + Z_0 }}.
\end{equation}
The  magnitude squared of the reflection coefficient is also referred as the reflectance.

\begin{figure}
	\includegraphics{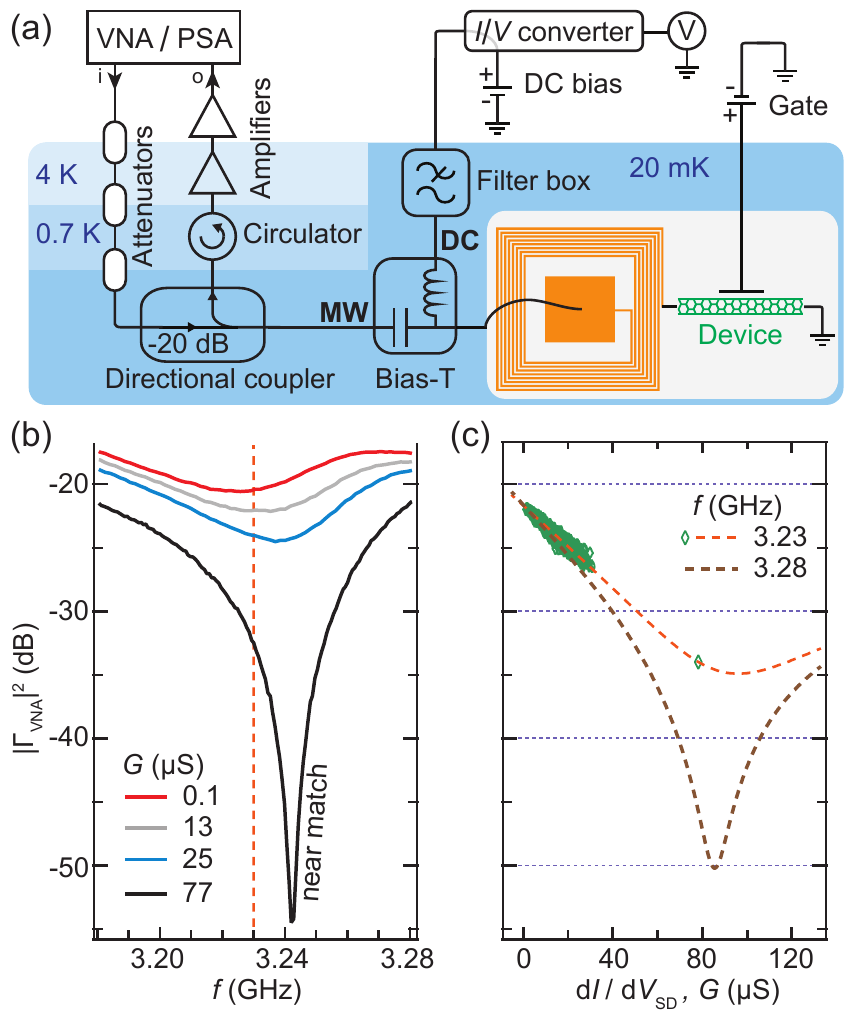}
	\caption{
		\label{fig2}
		(a) Schematic of the measurement setup with DC input and MW input (i), output (o) lines. The setup is used alternatively for reflectometry (both MW input and output) and noise measurements (only MW output), employing either a VNA or an PSA (vector network or power spectrum  analyzer).
		(b) Magnitude squared of the reflection coefficient $\Gamma_\reg{VNA} \equiv V_\reg{o}/V_\reg{i}$ measured around the resonance frequency. The further investigated regime lies between the close-to-zero conductance (Coulomb blockade) red curve (shallowest trace) and the vicinity of the blue curve. The separately measured black trace corresponds to an almost perfectly matched conductance.
		(c) Fit curve (orange) of the reflectance data $\left| \Gamma_{\rm VNA} \right| ^2 (\didv)$ collected at the fixed frequency 3.23 GHz during a $V_\reg G$ - $V_\reg{SD}$ 2D sweep. All the measured $\didv$ values (green diamonds) correspond to less than 30\,$\mu$S, except for one point taken from the near-match curve in (b). The deeper curve (brown) is calculated with the extracted fit parameters, at the extracted resonance frequency, and its minimum indicates $G_\reg{match}$.
	}
\end{figure}

The complete measurement setup, depicted in Fig.\,\ref{fig2}(a), is the same as in Ref.~\cite{thomas}. The load of the matching circuit, the CNT device is tuned by $V_\reg{SD}$ and $V_\reg{G}$---the DC source-drain and gate voltage, respectively. MW lines are used either in reflectometry mode (a microwave signal is sent to the sample and a reflected signal is read from it) or in noise mode where the MW output line conveys the noise signal produced by the sample. The two respective modes utilize a vector network analyzer (VNA) and a power spectrum analyzer (PSA).

We characterize the impedance-matching $LC$ resonator by analyzing reflectance measurements together with the simultaneously measured DC current $I$ from which we derive the differential conductance $\didv$. The $LC$ parameters we thus obtain allow us to extract the QD noise from the measured noise power, as we will show.

In practice, the measured reflection coefficient is not measured at the sample, but rather at the VNA, yielding $\Gamma_\reg{VNA} \equiv V_\reg{o}/V_\reg{i}$, where o and i refer to the VNA ports (see Fig.\,\ref{fig2}(a)). $\Gamma_\reg{VNA}$ is different from $\Gamma$ by a frequency-dependent baseline $b(f)$ produced by attenuations and amplifications in the setup:
\begin{equation}
  \label{baseline}
    |\Gamma_\reg{VNA}|^2 = b(f) \cdot |\Gamma|^2 .
  \end{equation}
If spurious reflections occur in components across the output line, then standing waves reside in sections of it and the baseline $b(f)$ can exhibit a complicated pattern.

In principle, we need to extract 5 parameters, $L$, $C$, $b$, $R_\reg{loss}$, and $R$. There are two possible approaches: one can fit the frequency dependence of $\Gamma_\reg{VNA}$ for a fixed known value of $G$, for example for $G=0$, which is realized when the QD is deep in Coulomb blockade where the current is almost completely suppressed. Fig.\,\ref{fig2}(b) shows three measured curves for different $G$ values. Unfortunately, the baseline function $b(f)$ is also markedly frequency-dependent, shifting the minimum of the curves to different values. Due to this dependence, it is already difficult to extract the resonance frequency accurately.

We therefore proceed with the second approach. We fix the frequency to a value close to resonance, in the following to $f_\reg{m}=3.23\,\reg{GHz}$.
We carry out simultaneous measurements of reflectance and DC current over several Coulomb diamonds. We obtain the reflectance $|\Gamma_\reg{VNA}|^2(V_\reg{G},V_\reg{SD})$ and the low-frequency $\didv(V_\reg{G},V_\reg{SD})$ maps. For each pair of $(V_\reg{G},V_\reg{SD})$ this defines one point in the $\Gamma_\reg{VNA}$-\,versus\,-\,$\didv$ scatter plot. To obtain the circuit parameters, we fit to this scatter plot the theoretical dependence $\Gamma_\reg{VNA} (G)$, thereby implicitly assuming that $\didv=G$. We have shown in previous work that the real part of the CNT admittance at similar GHz frequencies is the same as $dI/dV_\reg{SD}$ ~\cite{Ranjan2015,thomas}. The curve $\Gamma_\reg{VNA}(G)$ can be calculated using Eqs.\,(\ref{eq_Zin},\ref{eq_Gamma},\ref{baseline}). It is displayed in Fig.\,\ref{fig2}(c) for a wide range of $G$ values. At the resonance frequency, when sweeping the conductance domain from small to large $G$ values, $\Gamma_\reg{VNA}(G)$ first decreases to reach zero reflectance at the matching point where $G=G_\reg{match}$, then increases back for $G > G_\reg{match}$. The fit together with the measured scatter plot is shown in Fig.\,\ref{fig2}(c).

\section{Noise measurement calibration}
The current noise produced by a QD can be modeled by a noise voltage source $V_{\rm noise}$ as shown in Fig.\,\ref{fig1}(d,e). The power spectral density of the voltage source $S_V$ is related to current spectral density $S_I$ by $S_{V}=R^2 S_I$. The noise voltage needs to be transmitted through the $LC$ circuit to be fed into the $Z_0$\,-\,transmission line. The voltage $V$ that appears at the interface is proportional to $V_{\rm noise}$, but contains in addition a frequency-dependent transmission function $t_{V}(f) \equiv V/V_{\rm noise}$, which can readily be derived from the $LC$ circuit parameters:
\begin{equation}
	\label{eq_H}
	t_{V}(\reg \Omega) \approx \frac{Z_0 G}{1 + \reg{i}\Omega\left(\frac{Z_0}{Z_{\rm c}} + \frac{R_\reg{loss}}{Z_\reg{c}} + \frac{Z_\reg{c}}{R} \right) - \Omega^2},
\end{equation}
where $\Omega=\omega/\omega_0 = 2\pi f\sqrt{LC}$ and $Z_\reg{c}=\sqrt{L/C}$ the characteristic impedance as introduced before. In this equation, the factor of $\reg i \Omega$ shows that the total $Q$-factor has three terms, which can be grouped in an internal $1/Q_{\rm int}$ and an external loss part $1/Q_{\rm ext}$. The two parts are respectively $1/Q_{\rm int}=R_{\rm loss}/Z_\reg{c}$ and $1/Q_{\rm ext}=Z_0/Z_\reg{c} + Z_\reg{c}/R$. At optimal impedance matching in the lossless picture, i.e. if the condition $Z_\reg{c} = \sqrt{R Z_0}$ is met,
the external $Q$-factor is minimal and given by $\sqrt{R/Z_0}/2$ which yields values around $10-100$.

The transmitted voltage passes several elements, a directional coupler, a circulator and amplifiers at low and room temperature before being measured with the PSA. The overall gain along the chain is captured by $g$. It was determined with the method presented in Ref.\,\cite{supp_thomas} to be $g=94.6$\,dB.
In addition to the noise generated by the device, the amplifier does also add further current-independent noise. We split the total noise into two contributions $\langle \delta P\rangle$ and $\langle\delta P_{0}\rangle$, where the first term describes the noise power which is current-dependent and the second term contains all the rest, i.e. thermal and amplifier noise background. The different terms are related as follows~\cite{thomas}:
\begin{equation}
  \label{eq_si}
     \langle\delta P\rangle = \frac{S_I R^2}{Z_0} ~ g\int_{\rm BW}|t_V (f)|^2 {\rm d}f +\langle\delta P_0\rangle,
\end{equation}
where $S_{I}$ refers to the excess current noise only (i.e. the thermal noise of the device is included in $\langle \delta P_0 \rangle$).

\begin{figure}
  \includegraphics{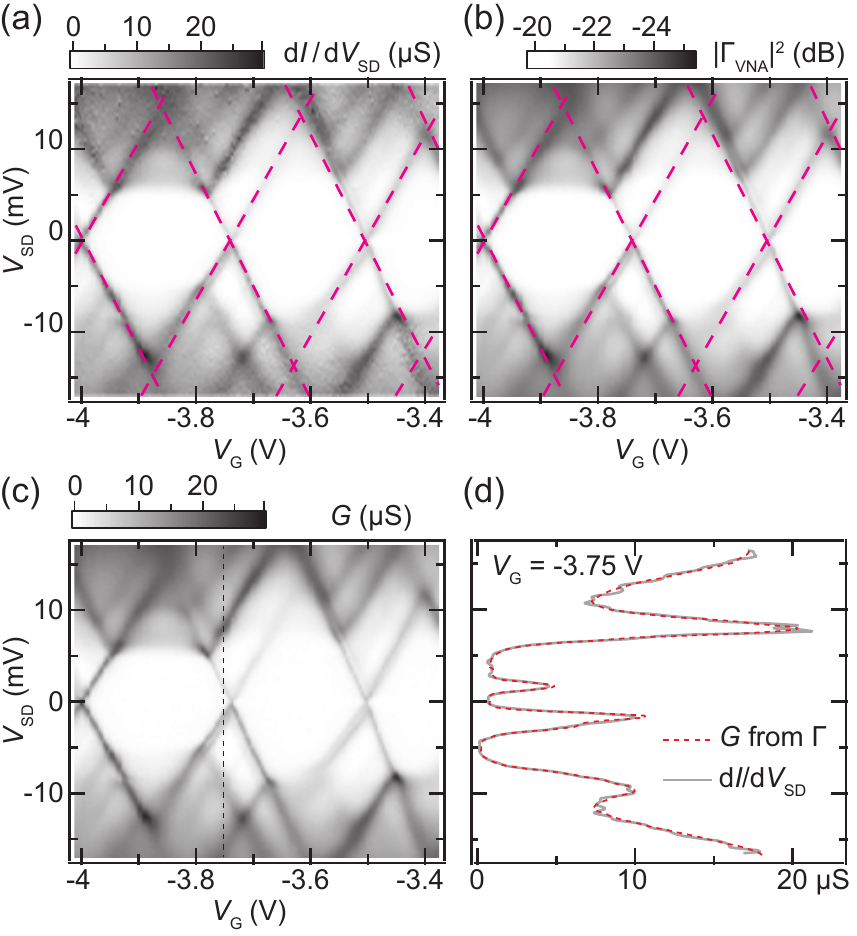}
    \caption{
	\label{fig3}
	  (a) Derivative of the DC current ($\didv$) as a function of the gate and source-drain voltages. The contour of the Coulomb diamonds is highlighted by the dashed line.	
	  (b) Reflectance measured at $f=3.23\,\reg{GHz}$, with the same contour as in (a).
	  (c) Differential conductance $G$ deduced from the reflectance $\left| \Gamma_\reg{VNA} \right| ^2 (f=3.23\,\reg{GHz},V_\reg{G},V_\reg{SD})$ by using the $LC$ matching circuit parameters extracted in fitting. (d) Cuts in the (a),(c) maps, showing an excellent overlap of low- and high-frequency differential-conductance values.
	}
\end{figure}

\section{Experiment}

By means of gate and bias voltage sweeps we measure the DC current and draw an initial charge stability diagram. Then we switch to MW reflectance measurements, where we restrict ourselves to a gate span covering a few Coulomb charge states. In this measurement mode, we simultaneously acquire the current $I$ and reflectance $|\Gamma_\reg{VNA}|^2$ as functions of $V_{\rm SD}$ and $V_{\rm G}$. The VNA was set to the working frequency $f_\reg{m}=3.23\,\reg{GHz}$ with
an output power of $-25\,\reg{dBm}$, attenuated by $\sm  87\,\reg{dB}$ before reaching the device.
We deduce the parameters of the $LC$ impedance-matching circuit from a comparison of the two measured maps $\didv(V_{\rm G},V_{\rm SD})$ (Fig.\,\ref{fig3}(a)) and $|\Gamma_{\rm VNA}(V_{\rm G},V_{\rm SD})|^2$ (Fig.\,\ref{fig3}(b)) as described in the previous section. In practice, the conductance domain of the resulted $|\Gamma_{\rm VNA}(\didv, f_\reg{m})|^2$ scatter plot does not include values close to $G_\reg{match}$, therefore we enrich the plot with a close-to-match point, $|\Gamma_\reg{VNA} ^\reg{near~match}(f_\reg{m})|^2$, from the separately obtained near-match curve in Fig.\,\ref{fig2}(b).

From complementary measurements we estimate an upper bound of $R_\reg{loss}\approx 1$\,$\Omega$ (see Supplemental material). Since $R_\reg{loss}\ll Z_0$, the matching circuit can be  modeled as a lossless $LC$ circuit for practical purposes.
Thus we fix $R_\reg{loss}=0$, and fit the curve $|\Gamma_\reg{VNA}|^2(G, f=f_\reg{m})$ (Fig.\,\ref{fig2}(c)), which yields $L=37$\,nH, $C=63$\,fF.
We further deduce the resonance frequency $f_0=3.28\,\reg{GHz}$ and the characteristic impedance $Z_\reg{c} =766\,\Omega$; the resulting match conductance is the minimum of the calculated curve $|\Gamma_\reg{VNA}|^2(G, f=f_0)$: $G_\reg{match}=84\,\rm {\mu}S$.
For comparison, refitting with $R_\reg{loss}= 1$\,$\Omega$, the calculated $L$, $C$, $G_\reg{match}$ vary by less than 1\,\%. 
Furthermore, for $R_\reg{loss}\leq1\,\Omega$ we obtain $Q_{\rm ext}\leq 15$ and $Q_{\rm in}\geq 766$, yielding
an empirical bandwidth $f_0/Q \gtrsim 220\,\rm MHz$.

Consequently, having fixed the parameters we then calculate a full map of the real part of the device admittance, i.e. $G(V_{G},V_{\rm SD})$ (Fig.\,\ref{fig3}(c)). A comparison of cuts (Fig.\,\ref{fig3}(d)) demonstrates a very good agreement with $\didv$, confirming the validity of the extracted parameters, at least at the frequency at which the reflectance map was acquired. The advantage of the reflectance measurement is, that $G$ can be measured much faster and with less noise, once the circuit parameters are known.

\begin{figure}
	\includegraphics{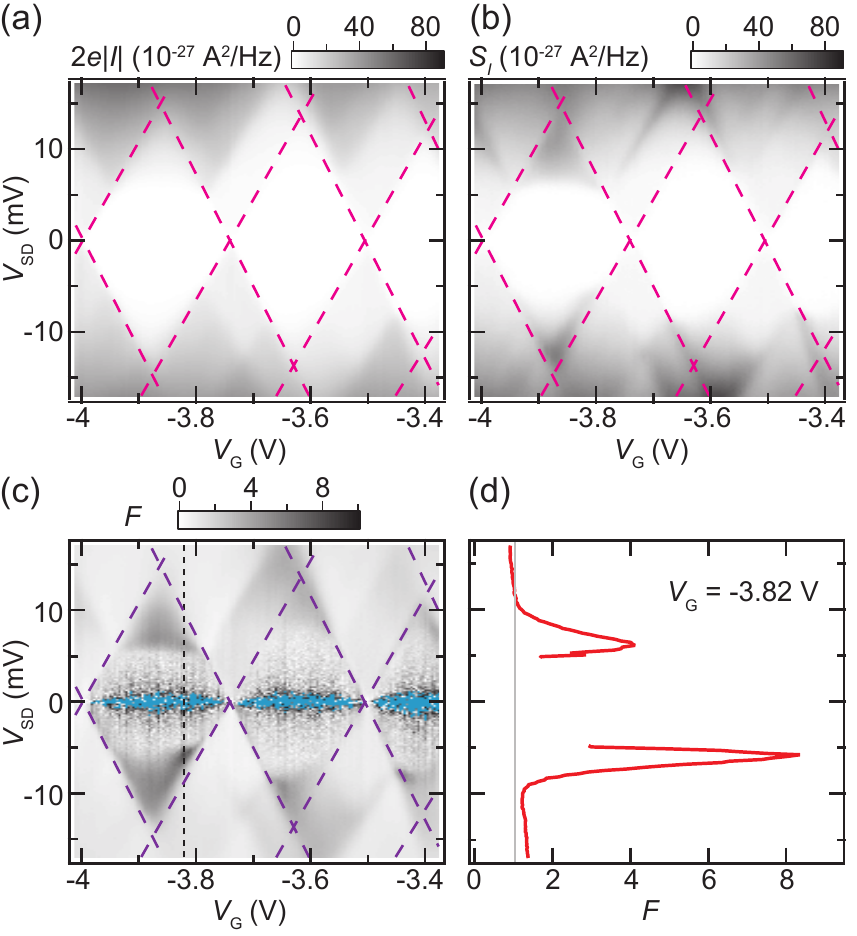}
	\caption{
		\label{fig4}
		(a) Shot noise spectral density $S_I$ as a function of $V_{\rm G}$ and $V_{\rm SD}$ obtained from the measured power-spectral density $\langle\delta P\rangle$ at the PSA using Eq.\,(\ref{eq_si}).
		(b) Schottky noise $2e|I|$ calculated from the DC measuered current $I$, and
		(c) Fano factor $F=S_I/ |2eI| $. The Coulomb diamond contours (dashed lines) are copied from the conductance plot in Fig.\,\ref{fig3}(a). Instead of $F=1$ proper to elastic cotunneling, the low-bias cyan regions indicate erroneously calculated values emerging from the division of a "noisy" number by a very small one. These regions need to be discarded. A cut at the black dashed vertical is depicted in (d). The observed highest Fano factor is $F \simeq 8$.
	}
\end{figure}

In the noise part of the experiment, we realize simultaneous current and noise measurements over the same few Coulomb diamond span as before. Each retained noise power value $\langle \delta P \rangle$ arises from averaging 500 identical-input measurements in a bandwidth $\reg{BW}=50\,\reg{MHz}$ and it is produced every $\sm$ 800\,ms.
Fig.\,\ref{fig4}(b) shows the current noise $S_I$ measured in the same gate range after applying Eq.\,\ref{eq_si} for calibration. To compare with the Schottky noise, we show in Fig.\,\ref{fig4}(a) a calculated plot of $2 e \lvert I \rvert$, where $I$ represents the measured DC current. The Fano factor, $F \equiv S_I / |2 e I |$, is shown in Fig.\,\ref{fig4}(c). We observe values largely exceeding 1 within the Coulomb blockade, where sequential tunneling of single electrons is forbidden. It is seen in Fig.\,\ref{fig4}(d) that the Fano factor peaks at $V_\reg{SD}\approx \pm 5$\,mV, where inelastic cotunneling sets in. Such a cotunneling event can excite the QD, enabling the sequential transfer of single electrons through the QD until the QD decays back to the ground state \cite{Wegewijs2001,Sukhorukov2001,Weymann2008,Gustavsson2008}.
The respective train of charge bunches leads to the large Fano factor, which here reaches values as large as 8. Only super-Poissonian noise with $F\leq 3$ has been reported earlier in QDs \cite{Onac2006,Okazaki2013}. At lower voltage bias, solely elastic cotunneling is present and the Fano factor should be 1 \cite{Sukhorukov2001,Okazaki2013}, this being the case in those regions of the second Coulomb diamond closer to the inelastic cotunneling onset; at even lower bias, the calculation of $F$ produces here divergent values, due to the division of "noisy" numbers by smallest current numbers, such that $F>10$ erroneously in the cyan domains of Fig.\,\ref{fig4}(c).

\section{Discussion and conclusion}
We have successfully  demonstrated the realization and application of a QD-impedance-matching on-chip superconducting coil working at 3.25\,GHz. We have deduced the $LC$ circuit parameters by using the dependence of the reflectance $|\Gamma|^2$ on the device resistance $R$ at a {\em fixed} frequency. The alternative approach, where one fits the frequency dependence of $|\Gamma|^2$ to the expected functional form for a fixed value $R$, typically at $G=1/R=0$ deep in the Coulomb-blockade regime, has turned out to be less reliable, due to the large bandwidth offered by the $LC$ circuit. Consequently, the circuit is affected by a frequency-dependent background $b(f)$ due to, for example, spurious standing waves in parts of the MW lines. As compared to transmission line resonators used as impedance-matching circuits, the figure of merit $g_\reg{SNR} = \reg{SNR/SNR_0}$, where $\reg{SNR}_0$ is the signal-to-noise ratio in the absence of any impedance-matching circuit~\cite{thomas}, has been shown to be the same for e.g. a stub tuner and an $LC$ circuit, if one acquires the noise over the full bandwidths given by the full widths at half maxima (FWHM) of the respective transfer functions $t_{\rm V}(f)$~\cite{supp_thomas}. The advantage of the $LC$ circuit lies in its larger bandwidth $\sm 100$\,MHz as compared to a transmission line resonator. This is important in application where short interaction times are essential, i.e. for fast readout and qubit manipulation. The disadvantage is that one has to account for spurious resonances in the external system leading to a frequency-dependent background. This can be a problem if highly accurate pulses need to be transmitted for qubit manipulation. Residual and uncontrolled phase shifts may make it difficult to achieve high gate fidelities. For noise measurement, we have shown it to be a powerful circuit.

\section*{Acknowledgements}
We acknowledge financial support from the ERC project QUEST and the Swiss National Science Foundation (SNF) through various grants, including NCCR-QSIT.

We thank Lukas Gubser for useful discussions.

\section*{Supplemental material}

\textbf{Loss characterization of a wire-bonded coil}

\begin{figure}
	\includegraphics{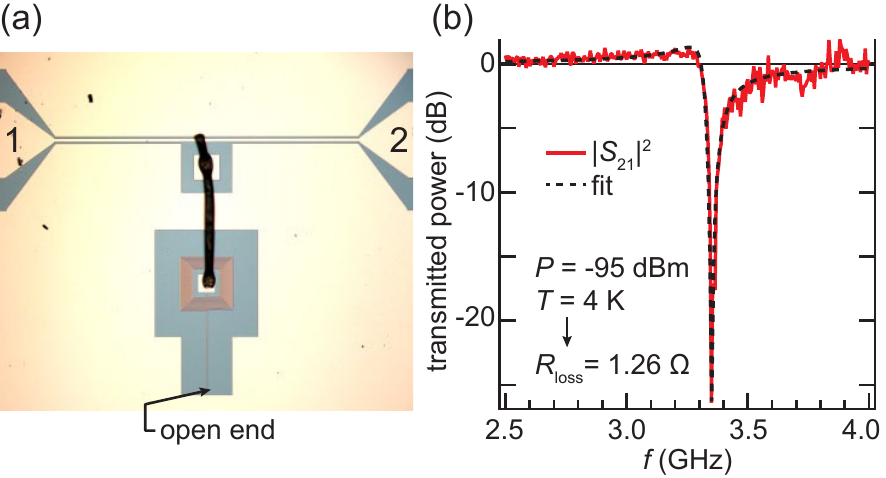} 
	\caption{
		\label{figS1}
		The loss resistance of a similar $LC$ circuit is extracted from (b) fitting $S_{21}$ of a device in a (a) hanger configuration with the coil connected to the middle of a transmission line; the bright color stands for metal and the bond wire is black.
	}
\end{figure}

We combined same-geometry coils and $50\,\Omega$ coplanar transmission lines in a ``hanger" configuration (Fig.\,\ref{figS1}(a)). We then measured at 4\,K the power transmission between ports 1 and 2 with $P = -95$\,dBm (higher than the excitation power used in the experiment).
We then fit the $S_{21}$ curve (Fig.\,\ref{figS1}(b)) with the formula \cite{Pozar2005}:

\begin{equation}
\label{eq_S21}
	S_{21} = \frac 2 {2+\frac {Z_\reg{line}} Z}, ~~Z_\reg{line} = 50\,\Omega \cdot e^{\reg i \phi}.
\end{equation}
Here, $Z=Z_\reg{in}(G=0)$ from Eq.\,(\ref{eq_Zin}) stands for a series $R_\reg{loss}LC$ cicuit; $\phi$ is a fitting parameter \cite{Khalil2012} that reproduces the asymmetry of the measured curve, caused by external spurious standing waves due to $50\,\Omega$ mismatches. The extracted parameters are: $R_{\reg{loss}} = 1.26\,\Omega$, $Z_\reg c = 954\,\Omega$, $f_\reg 0 = 3.35$\,GHz, $\phi = 0.56$.
The obtained loss resistance at 4\,K is a comfortable upper limit of our 20\,mK experiment, mainly because the Al bond wire becomes superconducting and Nb superconductivity gets reinforced in the mK range.

The reduced characteristic impedance in the main experiment $Z_{\rm c}= 766$\,$\Omega$ as compared to $Z_{\rm c}= 954$\,$\Omega$ in the hanger is explained by a decreased inductance from the two-turn short in the coil. However, the frequency (and $L\cdot C$) is very similar, suggesting that a short can also induce an increase in capacitance. The distributed structure of inductive and capacitive elements in a coil could indeed produce this effect.

\bigskip
\textbf{Fitting procedure}

\begin{figure}
	\includegraphics{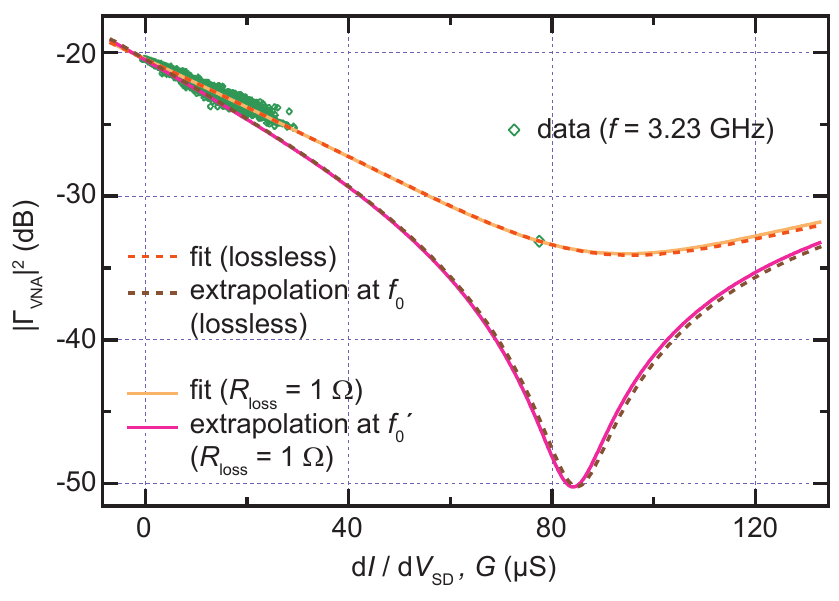} 
	\caption{
		\label{figS2}
		Reflectance vs. conductance. Green diamonds represent measured $|\Gamma_\reg{VNA}|^2$ vs. $\didv$. The dashed (full) lines are extracted $|\Gamma_\reg{VNA}|^2(G)$ curves at 3.23\,GHz  and at the resonance frequency, in the case $R_\reg{loss}=0$ ($R_\reg{loss}=1\,\Omega$).	}
\end{figure}

We extract the unknown circuit parameters $L$, $C$, $R_\reg{loss}$, and $b(f)$ by first fixing a frequency $f_m$ close to the resonance frequency. We then plot the measured reflectance values $|\Gamma_{\rm VNA}|^2$ against the DC measured $\didv$ for the same values of $V_\reg{G}$ and $V_{\rm SD}$. This is shown in Fig.\,\ref{figS2} as green diamonds. We next assume that $\didv$ is equal to $G$ at GHz frequency. Then we can fit a theoretical dependence to the data points, based on Eqs.\,(\ref{eq_Zin},\ref{eq_Gamma},\ref{baseline}). Both the dashed orange and full yellow trances in Fig.\,\ref{figS2} are candidate fitting curves.
The reflectance values at $G=0$ and $G \rightarrow \infty$ yield the baseline $b(f=f_\reg m)$ and $R_\reg{loss}$; the lack of the latter $G$ value explains why these two parameters cannot be reliably determined simultaneously. We therefore fix $R_\reg{loss}=0$; this choice is also a reasonable in further noise extraction, as $R_\reg{loss}$ is in series with the much larger characteristic impedance $Z_0=50\,\Omega$ of the setup output line.
One could be convinced of the low influence of $R_\reg{loss}$ by observing the overlap of the lossless and lossy fits in Fig.\,\ref{figS2}. In spite of the stability of the extracted parameters in the two fit cases (less than 1\,\% for $L$ and $C$), it is important to observe in Figs.\,\ref{fig2}(b),\,\ref{figS2} a $\sm 10\,\%$ deviation of the extracted $G_\reg{match}$ value from the measured one (77\,$\mu$S). Therefore the accuracy of the  extracted $S_I$ is at most $10\,\%$. In conclusion, working at a fixed frequency outputs effective values of $L$ and $C$ at that specific frequency, while the analysis of a whole frequency range, if possible, should provide more precise circuit parameters.

\end{document}